\documentclass[
  prd,
  reprint,
  nofootinbib,
  floatfix,
  aps
]{revtex4-2}

\usepackage{graphicx}
\usepackage{xcolor}

\usepackage{amsmath,amssymb}
\usepackage{multirow}

\usepackage{gensymb}      % provides \degree

\usepackage[normalem]{ulem} % for \sout; remove for final if possible
\usepackage{orcidlink}

\usepackage{hyperref}
\usepackage{cleveref}
\usepackage{xspace}

\newcommand{\nuwro}{\text{\sc NuWro}\xspace}

\newif\ifsupplement
\supplementtrue % <-- Ustaw \supplementtrue (jeden PDF) lub \supplementfalse (dwa PDF-y)
\ifsupplement
  \newcommand{\suppref}[1]{\ref{#1}}
  
\else
  \usepackage{xr-hyper}
  
  \newcommand{\suppref}[1]{\ref{supp:#1}}
  
\fi

\begin{document}

\newcommand{\luis}[1]{\textcolor{blue}{#1}}
\newcommand{\kg}[1]{\textcolor{red}{#1}}

%\title{GANs and flux}
\title{Machine-learning-based neutrino-event generation constrained by experimental measurements}

\author{Jos\'e L. Bonilla\,\orcidlink{0009-0009-3240-1494}}
\email{joseluis.bonillaramirez@uwr.edu.pl}

\author{Krzysztof M. Graczyk\,\orcidlink{0000-0002-0038-6340}}
\email{krzysztof.graczyk@uwr.edu.pl}

\author{\\Rwik Dharmapal Banerjee\,\orcidlink{0000-0003-3639-7532}}
\author{Beata E. Kowal\,\orcidlink{0000-0003-3646-1653}}
\author{Hemant Prasad\,\orcidlink{0009-0003-1897-9616}}
\author{Jan T. Sobczyk\,\orcidlink{0000-0003-4991-2919}}

\affiliation{Institute of Theoretical Physics, University of Wroc\l aw, plac Maxa Borna 9,
50-204, Wroc\l aw, Poland}

\date{\today}%

\begin{abstract}
We present a generative surrogate for charged-current \(\nu_\mu\)-nucleus scattering that can be calibrated to measured differential cross sections without event-level experimental samples. Starting from a theoretical interaction model, we fine-tune the surrogate to a MicroBooNE measurement and show that this also improves the description of observables not used in the fit. The calibrated model can also be applied to measurements from different neutrino fluxes without retuning, while retaining event-level predictions of outgoing-muon kinematics. Our results show that experimental cross sections can directly constrain generative event models while preserving their applicability across different observables and beam configurations.
\end{abstract}

\maketitle

\normalem

\textit{Introduction.} Precision measurements of neutrino oscillations and leptonic CP violation are demanding tests for establishing the Standard Model's consistency. They may also shed light on the matter-antimatter asymmetry of the Universe \cite{Fukugita:1986hr}. To achieve these precision measurements at next-generation long-baseline neutrino experiments, such as DUNE~\cite{DUNE:2020lwj} and Hyper-Kamiokande~\cite{Hyper-KamiokandeProto-:2015xww}, an accurate modeling of neutrino–nucleus interactions is required. Nuclear effects matter because they affect final states and neutrino-energy reconstruction~\cite{Martini:2012fa,Mosel:2013fxa}. Recent studies show that uncertainties in final-state interaction modeling can induce distortions comparable to the precision required for oscillation-parameter measurements~\cite{Liu:2025hpl}. At the same time, increasingly sophisticated theoretical neutrino interaction models make efficient event generation in Monte Carlo (MC) simulation tools increasingly challenging~\cite{Andreopoulos:2009rq,Golan:2012wx,NuSTEC:2017hzk,Mosel:2018qmv,Hayato:2021heg,Isaacson:2022cwh,campbell2024eventgeneratorshighenergyphysics}. Generative deep-learning models provide natural surrogates for stochastic particle–nucleus scattering, learning multidimensional distributions of final-state observables and allowing efficient event sampling~\cite{Pina-Otey:2020hzm,ElBaz:2025qjp,Alanazi:2020jod,bonilla2025generativeadversarialneuralnetworks,Bonilla:2025rrn}.
Beyond reproducing a single process, transfer learning can extend such models to related processes~\cite{graczyk2024electronnucleuscrosssectionstransfer}.

Accelerator neutrino beams are not monoenergetic but have a broad, energy-dependent flux~\cite{T2K:2019bcf}. Measured cross sections are flux-averaged, with several reaction mechanisms contributing simultaneously, complicating both model validation and MC event-generator tuning. For a fixed nuclear target, however, interaction physics, nuclear effects, and cross section distributions depend on $(E_\nu,\text{flavor})$, independently of how the neutrinos were produced. 
Detector response, acceptance, and backgrounds are experiment-specific and enter the measurement and its uncertainties. At the interaction level, however, experiments using the same nuclear target differ primarily through their neutrino flux \(\Phi(E_\nu,\mathrm{flavor})\).
A surrogate conditioned on $E_\nu$ can therefore transfer interaction information between beam configurations, with the beam profile as the experiment-specific input.

We realize this with a generative adversarial network (GAN) which readily learns unnormalized event
distributions for neutrino and antineutrino scattering off
nuclei~\cite{bonilla2025generativeadversarialneuralnetworks,Bonilla:2025rrn}, but an unnormalized
shape cannot be compared to a measured cross section nor transported to a different beam. The missing ingredient is the energy-dependent interaction probability. 
This weighting is precisely what a bare generator lacks. We therefore couple the GAN to a deep neural network (DNN) that predicts the total cross section $\sigma(E_\nu)$, so we can normalize generated events to the cross section. The surrogate can then be fine-tuned directly on flux-averaged measurements, and its interaction-level predictions can be transported to other experiments by supplying the corresponding neutrino flux, enabling a direct comparison between the surrogate
and measurements not used in the training. 
In this sense, the framework provides a way to fine-tune an MC interaction model directly from the measured cross sections, without requiring event-level experimental information. This differs from machine-learning unfolding approaches such as OmniFold~\cite{Andreassen:2019cjw}, which iteratively reweight simulated events using detector-level data and detector simulation. Instead, our approach modifies the generative model and is constrained directly by binned cross sections and their covariance, rather than per-event detector-level samples. It also differs from the parametric generator tunes~\cite{MicroBooNE:2021ccs}, which adjust a predefined set of parameters of the underlying interaction model.

The central idea is to retain the event-level structure learned from a
physics-based MC model while allowing measured cross sections to modify
its predictions. We therefore couple a $\sigma$-DNN, which supplies the
energy-dependent normalization, to a conditional GAN that generates the
outgoing-muon kinematics $(E_\mu,\theta)$ at fixed $E_\nu$. Relative to
Ref.~\cite{bonilla2025generativeadversarialneuralnetworks}, the key extensions are the explicit normalization for
arbitrary neutrino fluxes and direct fine-tuning to measured cross
sections rather than simulated events alone. The resulting surrogate
therefore combines MC-based event structure with direct experimental
constraints and can be applied across different beam configurations.

We test this strategy by calibrating the model to a MicroBooNE measurement~\cite{MicroBooNE:2023foc} of the inclusive charged-current (CC) $\nu_\mu$-argon scattering double-differential cross section $d^2\sigma/dp_\mu d\cos\theta$, and evaluating it against measurements involving different observables and neutrino fluxes. The tuned surrogate improves the description of MicroBooNE observables not used in the training~\cite{MicroBooNE:2021sfa,MicroBooNE:2024zkh,MicroBooNE:2024zwf} and is consistent with the ArgoNeuT measurements~\cite{ArgoNeuT:2011bms, ArgoNeuT:2014rlj} obtained with a substantially different flux. These results show that experimentally measured cross sections can constrain a generative event model while remaining applicable across different observables and beam configurations.

\begin{figure}
\includegraphics[width=0.48\textwidth]{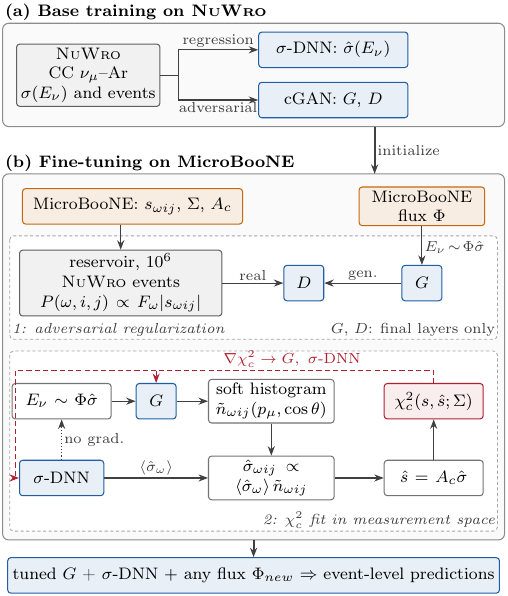}
\caption{Training scheme: (a) The \(\sigma\)-DNN and GAN are pretrained on NuWro. (b) Fine-tuning alternates adversarial regularization using the data-informed reservoir with a differentiable full-covariance fit to the forward-smeared MicroBooNE measurement. The latter updates \(G\) and the \(\sigma\)-DNN. \label{fig:scheme_main}}
\end{figure}

We construct our model in two stages. First, we obtain the $\sigma$-DNN and GAN models independently from data produced with the \nuwro{} MC event generator. Then we fine-tune both models simultaneously using experimental data. Fig.~\suppref{fig:scheme_main} illustrates the model optimization scheme.

\textit{The first stage.} The $\sigma$-DNN is a feed-forward fully connected network. Rather than mapping a single energy to a cross section, it predicts the cross section averaged over an energy interval, with the single-energy value recovered for a vanishing interval width.

Using the $\sigma$-DNN and a known experimental flux $\Phi$ (in our case, from  MicroBooNE), we obtain a sample of interacting neutrino energies. Namely, the total-cross-section-weighted energy distribution of incoming neutrino events reads:
$f_i = {\hat\sigma_i\Phi_i}/{\langle\hat\sigma_\Omega\rangle}$,
where $\hat\sigma_i$ is the total cross section predicted by the $\sigma$-DNN (or by \nuwro{}) (averaged over the $i$-th bin), and 
$\langle\hat\sigma_\Omega\rangle = \sum_{i\in\Omega} \hat\sigma_i\Phi_i$,
is the flux-averaged cross section, where the flux satisfies:
$\sum_{i\in\Omega} \Phi_i = 1$, and $\Omega$ is the domain covered by the flux. 

To generate events, we use a GAN conditioned on the neutrino energy $E_\nu$ to reproduce the muon kinematics as a function of $E_\nu$. The generator $G$ maps a latent noise vector, together with $E_\nu$, to the muon energy and scattering angle $(E_\mu, \theta)$; the discriminator $D$ learns to distinguish these generated events from \nuwro{} training events. We trained the network from scratch on CC $\nu_\mu$-argon events generated with \nuwro{} in the same manner as in Ref.~\cite{Bonilla:2025rrn}. Technical details of the \(\sigma\)-DNN, GAN architecture and training, fine-tuning procedure, differentiable histogramming, and additional comparisons are given in the Supplemental Material (SM)~\cite{SupplementalMaterial}.

Using the event distribution $f_i$, we randomly
sample neutrino energies for the GAN generator to obtain
cross-section-weighted event samples. Differential cross sections are then
reconstructed from the generated event fractions and the corresponding
$\sigma$-DNN normalization; the explicit construction used in the
fine-tuning is given in the SM.

\textit{In the second stage}, we fine-tune the $\sigma$-DNN and GAN to the
MicroBooNE CC $\nu_\mu$-argon measurement
$d^2\sigma/dp_\mu\,d\cos\theta$~\cite{MicroBooNE:2023foc}.
We achieve this by fine-tuning the models using a nonstandard procedure. The difficulty is twofold:
the MicroBooNE data are presented as a three-dimensional differential
cross-section distribution (double-differential in bins of $E_\nu$), whereas the GAN is trained on event samples; and the
differential distributions also encode the total cross section, which in our
system is reproduced by the $\sigma$-DNN. Both observations lead us to a joint optimization procedure in which the GAN and $\sigma$-DNN weights are updated together
using the experimental distributions.

The second-stage optimization combines two complementary sub-steps. \textit{In the first}, the discriminator is fine-tuned on events drawn from  a reservoir, while the generator is updated adversarially against the discriminator; only the final layers of both networks are allowed to adapt. The generator samples neutrino energies from the flux reweighted by the current \(\sigma\)-DNN prediction.

The reservoir consists of one million truth-level \nuwro\ events resampled
from a larger sample generated with the MicroBooNE flux. The MicroBooNE
result is a smeared double-differential cross section in \((p_\mu,\cos\theta_\mu)\), reported in bins of \(E_\nu\), with the smearing matrix $A_c$ relating
the truth-level prediction to the measured bin space. The measured bin
contents, together with the MicroBooNE flux, determine non-negative target
occupancies for the reservoir, as detailed in the SM. Since the reservoir
is constructed from truth-level \nuwro\ events using a smeared measurement,
it cannot be interpreted as a truth-level target; rather, it regularizes
the generator toward the physical, \nuwro-like manifold. We perform the
quantitative fit separately in the second sub-step, where we forward-smear
the generated prediction with $A_c$ and compare it with the original
measurement.

\textit{In the second sub-step}, we update the generator and the $\sigma$-DNN jointly using the differential cross-section measurement. We histogram events generated by $G$, and normalize each histogram by the $\sigma$-DNN prediction for the flux-averaged cross section $\langle\hat\sigma_\omega\rangle$ over the corresponding neutrino-energy interval $\omega$. This yields an estimate of the double-differential cross section in each $E_\nu$ interval. The prediction is then forward-smeared,
$\hat s_\omega=A_c\hat\sigma_\omega$, and compared with the measured
$s_\omega$ through a $\chi_c^2$ constructed with the full covariance matrix;
see Eq.~\ref{eq:chic_app} of the SM.

The full covariance is retained because the unfolding and the smearing $A_c$
induce bin-to-bin correlations. If we do not follow this prescription, the fitting procedure collapses, as discussed in Ref.~\cite{Abe:2025yaa}. To make the
comparison differentiable with respect to the network weights, the histogramming
step is replaced by a smooth approximation, detailed in
Sec.~\suppref{app:secondstep} of the SM. The sampling of
$E_\nu$ conditions the generation but does not contribute to the gradient; the
normalization through $\langle\hat\sigma_\omega\rangle$ does.

\begin{table}[htbp]
    \caption{The values of $\chi^2_v$ and $\chi^2_c$ over the number of bins (n) for four configurations of the models: base (subscript B) and fine-tuned (subscript T) $\sigma$-DNN and GAN models. Computed for samples of 1 million events. \label{tab:chi2metrics}}%The reference values are obtained for \nuwro\ 21.09.
    \begin{ruledtabular}
        \begin{tabular}{lcc}
                                                              & $\chi_v^2$/n & $\chi_c^2$/n \\ \hline%\hline
            %\nuwro\ 21.09                                 & 460.8/138      & 449.3/138      \\ %\hline
            % chi2_c training
                $\sigma$-DNN$_\text{B}$, GAN$_\text{B}$ & 461.7/138      & 460.3/138      \\ %\hline
                $\sigma$-DNN$_\text{B}$, GAN$_\text{T}$ & 435.6/138      & 268.4/138      \\ %\hline
                $\sigma$-DNN$_\text{T}$, GAN$_\text{B}$ & 152.0/138      & 500.1/138      \\ %\hline
                $\sigma$-DNN$_\text{T}$, GAN$_\text{T}$ & 110.5/138      & 220.3/138      \\ %\hline
        \end{tabular}
    \end{ruledtabular}
\end{table}
\begin{figure}[htbp]
    \includegraphics[width=0.38\textwidth]{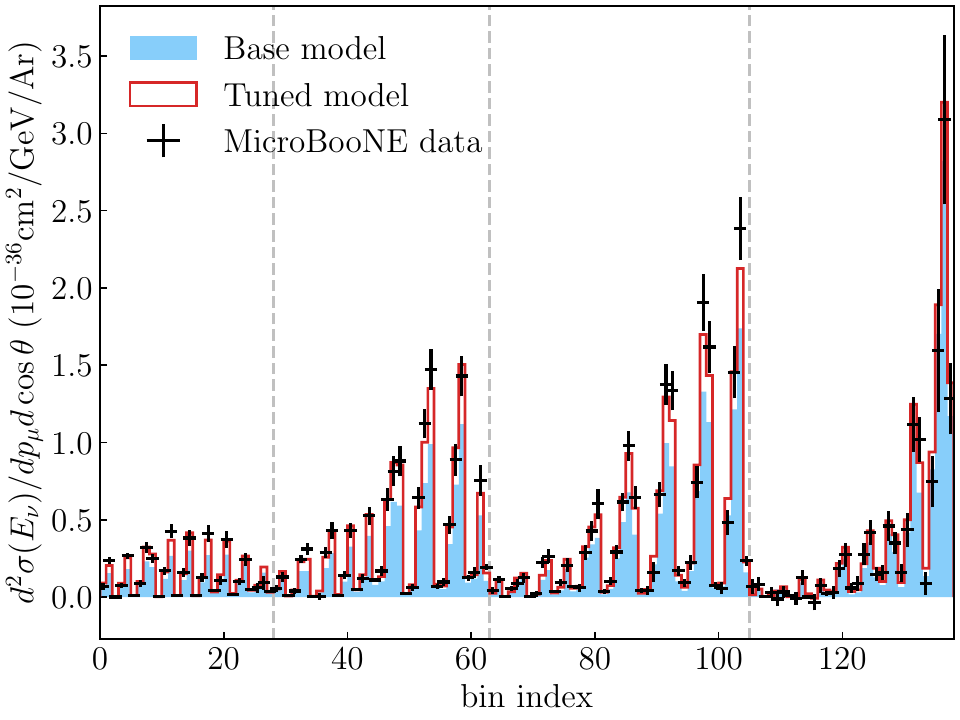}
    \caption{Base and tuned models compared with the MicroBooNE double-differential cross-section measurement~\cite{MicroBooNE:2023foc} for various \(E_\nu\) ranges. Grey vertical dashed lines separate different $E_\nu$ regions, from left to right: $[0.2,0.705]$, $[0.705,1.05]$, $[1.05,1.57]$, and $[1.57,4]$ GeV.\label{fig:histstune}}
\end{figure}

Table~\suppref{tab:chi2metrics} summarizes the effectiveness of the fine-tuning,
while Fig.~\suppref{fig:histstune} illustrates the resulting agreement with the
data. We report two complementary metrics: the diagonal $\chi^2_v$, defined in
Eq.~(\suppref{eq:chiv_main}) of the SM, which primarily
probes per-bin agreement, and the full-covariance $\chi^2_c$, which also
accounts for the bin-to-bin correlations induced by the unfolding and the
smearing matrix $A_c$. The table shows these metrics for the four possible
post-training combinations of the base and tuned $\sigma$-DNN and GAN
components.
\begin{figure}[htbp]
    \includegraphics[width=0.38\textwidth]{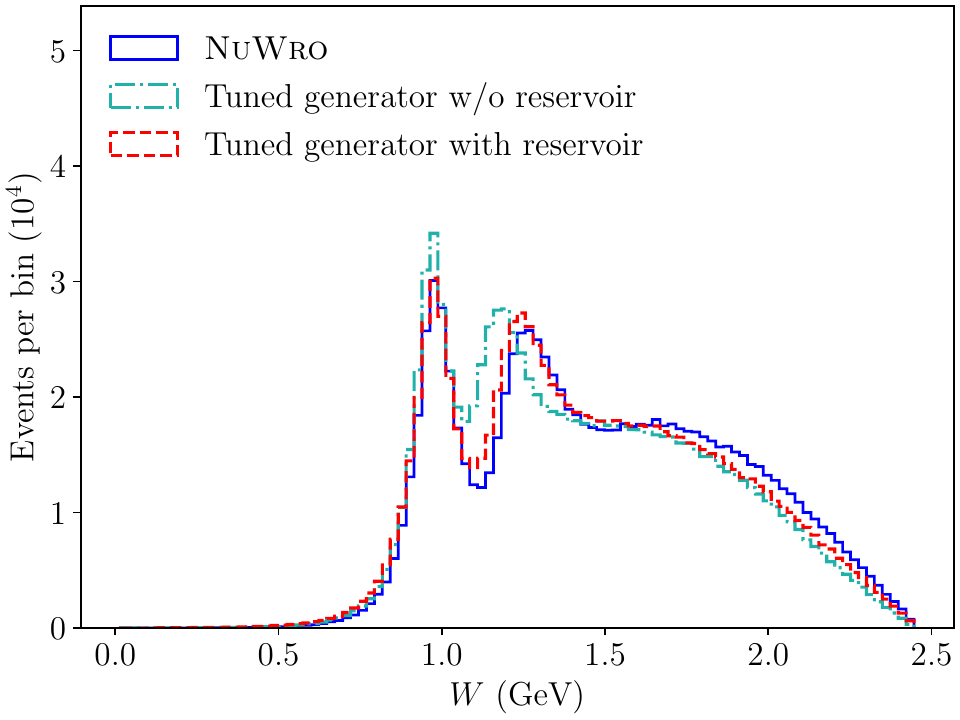}
    \caption{Event distribution at \(E_\nu=3\) GeV generated by the fine-tuned GAN with and without the reservoir sub-step.\label{fig:nochi2}}
\end{figure}

Replacing only the base $\sigma$-DNN with its tuned counterpart lowers
$\chi^2_v$ sharply but does not improve $\chi^2_c$, indicating that this component mainly corrects the
cross-section normalization. Conversely, replacing only the base GAN with the
tuned generator substantially improves $\chi^2_c$ while leaving $\chi^2_v$ nearly unchanged, showing its dominant role in reproducing correlated
kinematic structure. Only the combination of both tuned components improves
both metrics simultaneously, demonstrating their complementary roles and
motivating their joint optimization.
\begin{figure*}[htbp]
    \includegraphics[width=0.28\textwidth]{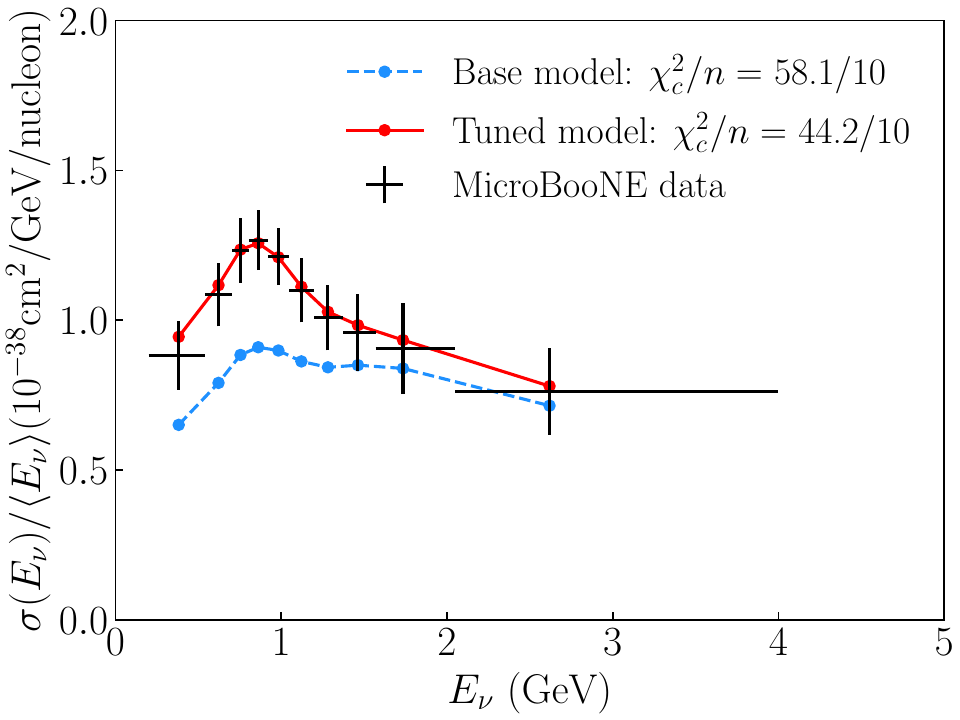}
    \includegraphics[width=0.28\textwidth]{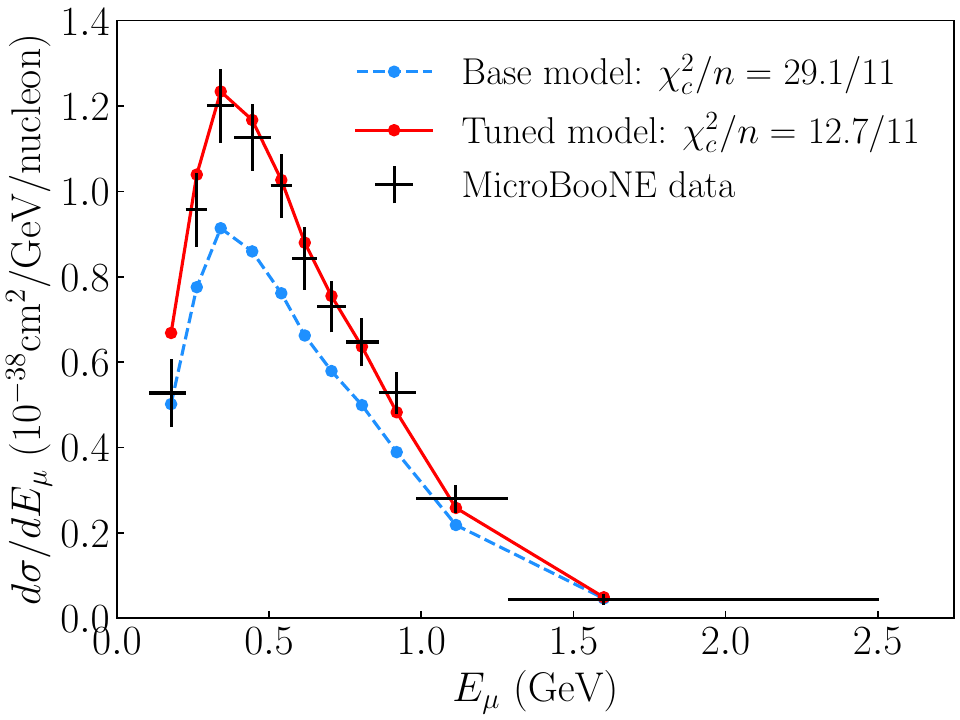}
    \includegraphics[width=0.28\textwidth]{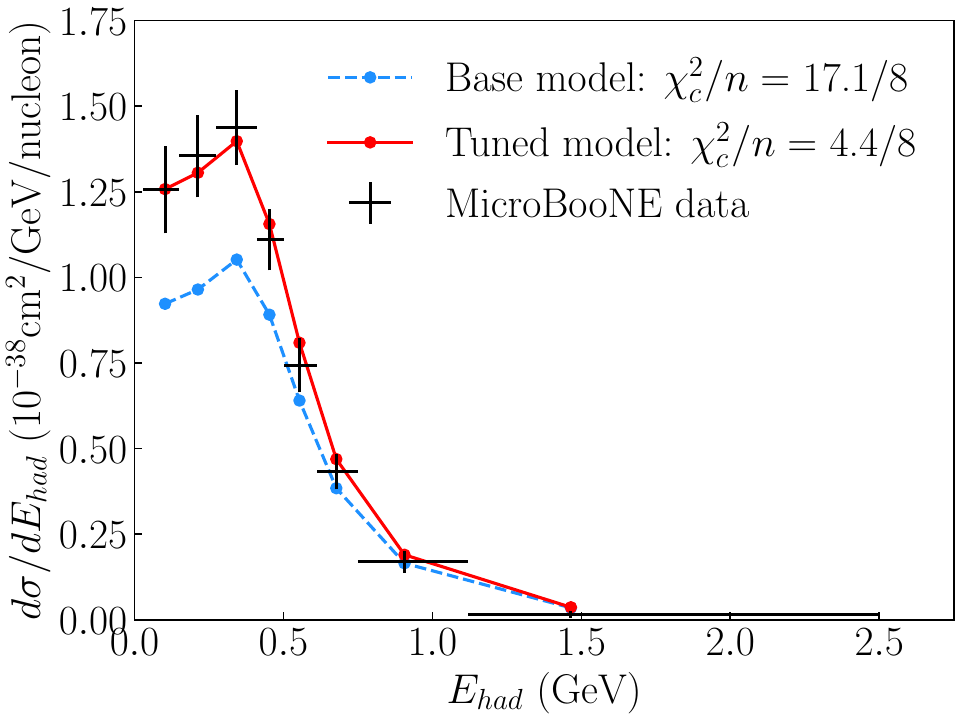}
    \caption{Base and fine-tuned models compared to the MicroBooNE measurement~\cite{MicroBooNE:2021sfa}: $\sigma/\langle E_\nu\rangle$ (left), $d\sigma/dE_\mu$ (center) and $d\sigma/dE_\text{had}$ (right).\label{fig:sigma_enu}}
\end{figure*}

We note that the first sub-step, in which the discriminator is exposed to the
reservoir, is essential. Without it, training with the $\chi^2_c$ loss
term alone reproduces the smeared, flux-averaged distributions but distorts the quasielastic (QE) and $\Delta$-resonance peaks; see Fig.~\suppref{fig:nochi2}: the
smearing $A_c$ and the coarse measurement binning wash out this sharp structure
in the observable to which $\chi^2_c$ is fit, so the covariance objective alone
cannot constrain it. The reservoir, built from truth-level \nuwro\ events that
carry this structure, supplies exactly the differential information the smeared
$\chi^2_c$ lacks, keeping the generator on the physical manifold of reaction
events.

We next test whether the tuned surrogate is consistent with other
published MicroBooNE cross-section measurements not used in the fit.
We first consider an earlier MicroBooNE measurement
\cite{MicroBooNE:2021sfa} of the total cross section $\sigma(E_\nu)$
and the differential cross sections $d\sigma/dE_\mu$ and
$d\sigma/dE_\mathrm{had}$, with
$E_\mathrm{had}=E_\nu-E_\mu$.
The model was tuned only on the cross section of
Ref.~\cite{MicroBooNE:2023foc}; the observables considered here are
therefore an out-of-training test of what the surrogate has learned,
not a refit.
Fig.~\suppref{fig:sigma_enu}
shows the corresponding cross sections. 
The tuned model improves the description of all three observables relative
to the base model (both in the shape and in the normalization):
in the full-covariance metric
$\chi^2_c/\text{n}$ improves from $58.1/10$ to $44.2/10$ for $\sigma(E_\nu)$,
from $29.1/11$ to $12.7/11$ for $d\sigma/dE_\mu$, and from $17.1/8$ to $4.4/8$
for $d\sigma/dE_\text{had}$. The improvement is most pronounced at neutrino
energies below $2$~GeV and muon energies below $1$~GeV, the region best
constrained by the data.

\begin{figure}[htbp]
    \includegraphics[width=0.38\textwidth]{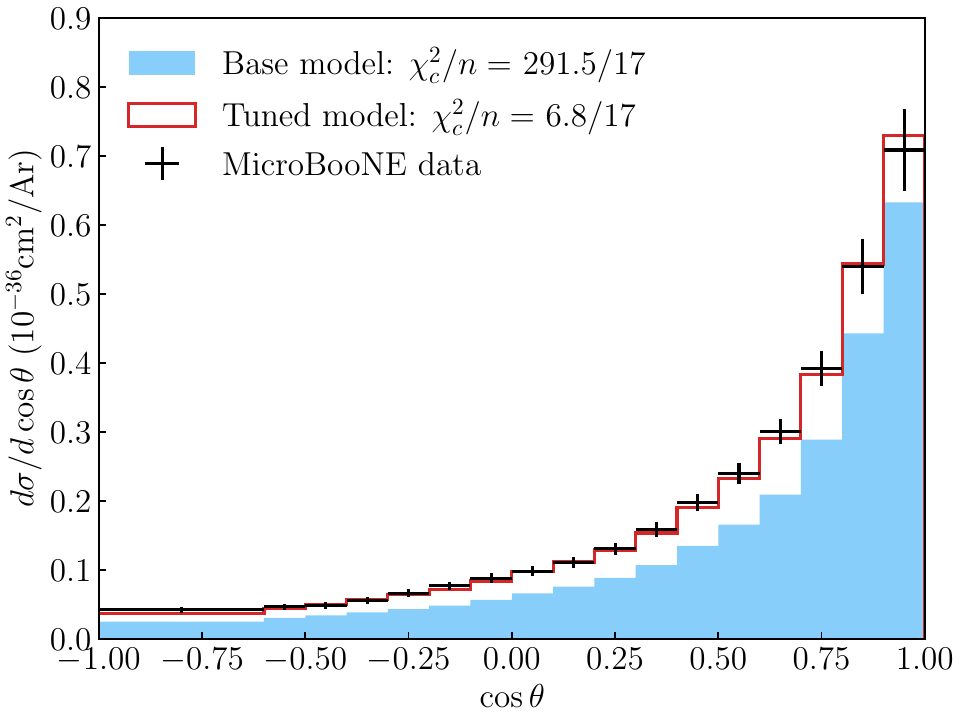}
    \caption{Base model, tuned model, and the MicroBooNE data~\cite{MicroBooNE:2024zkh,MicroBooNE:2024zwf} for $d\sigma/d\cos\theta$, not used in the training.\label{fig:double}
    }
\end{figure}

Finally, we compare our model against MicroBooNE and ArgoNeuT data not used in training. We start from more recent MicroBooNE
measurements~\cite{MicroBooNE:2024zkh,MicroBooNE:2024zwf} of 
$d\sigma/d\cos\theta$ cross section shown in Fig.~\suppref{fig:double}.
The comparison for the double-differential cross section $d^2\sigma/dE_\mu\,d\cos\theta$ ~\cite{MicroBooNE:2024zkh,MicroBooNE:2024zwf} is shown in SM. These measurements were not used at any stage of training or model selection; they therefore constitute a cross-measurement test of the surrogate. The tuned model improves markedly over the
base model in both observables. In the full-covariance metric,
$\chi^2_c/\text{n}$ drops from $257.0/69$ to $109.2/69$ for
$d^2\sigma/dE_\mu\,d\cos\theta$, and from $291.5/17$ to $6.8/17$ for
$d\sigma/d\cos\theta$. The latter is a notable result: a model tuned on a single,
distinct measurement predicts a later measurement not used in training at $\chi^2_c/\text{n}<1$,
without any retuning. This shows that the data-driven adaptation remains consistent with a different representation of the same underlying interaction kinematics.

\begin{figure}[htbp]
    \includegraphics[width=0.38\textwidth]{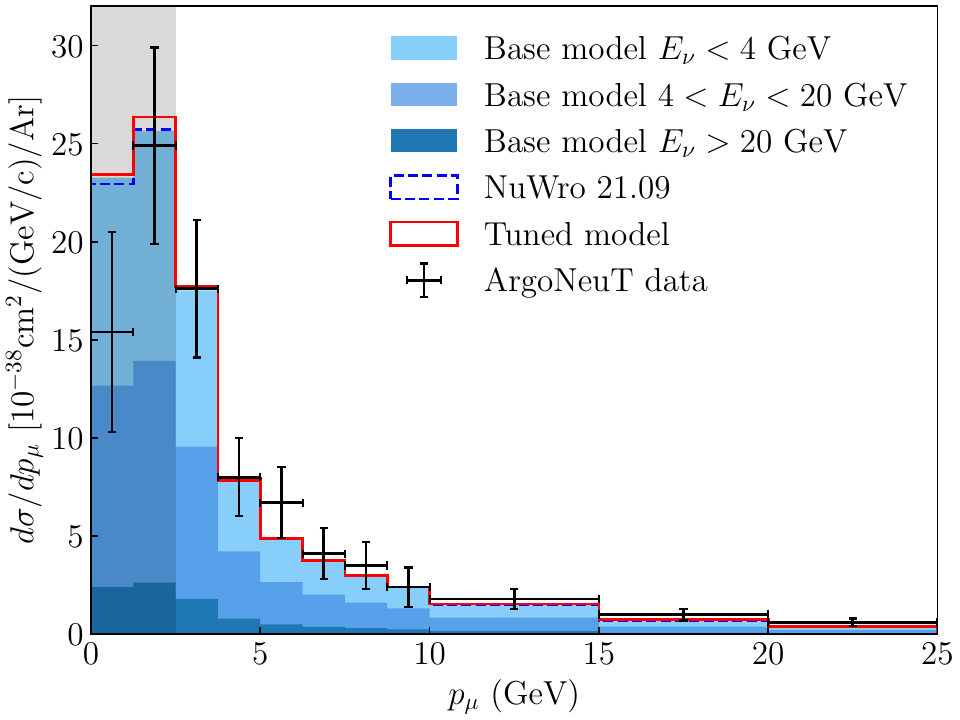}
    \caption{Base model, tuned model, and ArgoNeuT data~\cite{ArgoNeuT:2011bms, ArgoNeuT:2014rlj} for $d\sigma/dp_\mu$ distribution. The shaded area shows the $p_\mu$ range of the training MicroBooNE data.\label{fig:numi}}
\end{figure}

ArgoNeuT provides a qualitatively different test: it uses the NuMI beam
in its low-energy configuration~\cite{Adamson:2015dkw}, whose flux
extends well above the energies covered by the training data, and it is
a distinct detector. We therefore use it primarily to probe the
extrapolation of the framework rather than the data-driven adaptation
within the constrained region. We obtain predictions with \nuwro\ and
our DNN using the fluxes of
Refs.~\cite{ArgoNeuT:2011bms,ArgoNeuT:2014rlj}. ArgoNeuT did not
publish the covariance matrix, so a full $\chi^2_c$ cannot be formed;
we report the diagonal $\chi^2_v$ only as an indicator, noting that it
neglects the bin-to-bin correlations of the measurement.

Fig.~\suppref{fig:numi} shows the $d\sigma/dp_\mu$ distribution for
ArgoNeuT; see also the SM for the antineutrino-mode results. Because
the ArgoNeuT flux lies almost entirely above the energy range
constrained by the MicroBooNE fine-tuning, the tuned and base models
give similar flux-averaged predictions, and the agreement with the
data is therefore inherited mainly from the base model's high-energy
behavior. Both models yield
$\chi_v^2/\mathrm{n}\lesssim1$, although this partly reflects the large
ArgoNeuT uncertainties. Within these limitations, the tuned surrogate
remains consistent with the out-of-training ArgoNeuT data, indicating
that the fine-tuning does not substantially degrade the high-energy
behavior inherited from the base model.

\textit{In summary}, we have constructed a machine-learning-based neutrino-event generator that can be calibrated directly to binned experimental cross sections without event-level experimental samples. A conditional GAN generates the outgoing-muon kinematics (energy $E_\mu$ and scattering angle $\theta$), while a \(\sigma\)-DNN provides the energy-dependent absolute normalization; both are jointly constrained by the forward-smeared MicroBooNE measurement. The central result is that experimental data can update a generator pretrained on a conventional Monte Carlo model, and it can then be used for different observables and flux configurations.

A model tuned to the MicroBooNE measurement improves the description of other MicroBooNE observables not used in training and remains consistent with ArgoNeuT data obtained with a substantially different flux. The data-constrained region modifies the normalization and correlated kinematics, while outside this region the prediction largely inherits the underlying MC model. The framework therefore defines a route toward event generators that combine theory-based simulations with direct experimental constraints; extending the constrained kinematic range and propagating calibrated model uncertainties are natural next steps. With the latter in place, the surrogate could serve as a common interaction reference that, folded with each beam’s flux, tests the mutual consistency of cross-section measurements across experiments.

The fine-tuned models and the event-generation script are available in Ref.~\cite{author2026repo}.

\begin{acknowledgments}
This research is partly (K.M.G., B.E.K, J.T.S.) or fully (J.L.B, H.P., R.D.B.) supported by the Na{\-}tional Science Centre under grant UMO-2021/41/B/ST2/02778. J.L.B. and J.T.S. are also supported by the Polish Ministry of Science Grant No. 2022/WK/15. 
\end{acknowledgments}

% --- KONIEC GŁÓWNEGO ARTYKUŁU ---

%==================================================
% SUPPLEMENTAL MATERIAL
%==================================================

\ifsupplement

\clearpage
\onecolumngrid

\setcounter{section}{0}
\setcounter{subsection}{0}
\setcounter{subsubsection}{0}
\setcounter{equation}{0}
\setcounter{figure}{0}
\setcounter{table}{0}
\setcounter{page}{1}

% Printed numbering
\renewcommand{\thesection}{S-\Roman{section}}
\renewcommand{\theequation}{\thesection.\arabic{equation}}
\renewcommand{\thefigure}{S\arabic{figure}}
\renewcommand{\thetable}{S\arabic{table}}
\renewcommand{\thepage}{S\arabic{page}}

% Unique hyperlink destinations
\renewcommand{\theHsection}{supp.\Roman{section}}
\renewcommand{\theHsubsection}{supp.\Roman{section}.\arabic{subsection}}
\renewcommand{\theHsubsubsection}
  {supp.\Roman{section}.\arabic{subsection}.\arabic{subsubsection}}
\renewcommand{\theHequation}
  {supp.\Roman{section}.\arabic{equation}}
\renewcommand{\theHfigure}{supp.\arabic{figure}}
\renewcommand{\theHtable}{supp.\arabic{table}}

\makeatletter
\@addtoreset{equation}{section}
\makeatother

\begin{center}
    \textbf{\Large Supplemental Material}
\end{center}
\vspace{0.5cm}

\section{Model optimization}

In the following sections, we describe in detail all steps of the model optimization. 
Fig.~\ref{fig:scheme} presents an expanded version of the model-optimization scheme shown in Fig.~\ref{fig:scheme_main} of the main text.

\section{Cross-section DNN: parametrization and training}
\label{app:sigmadnn}

\begin{figure*}
\includegraphics[width=\textwidth]{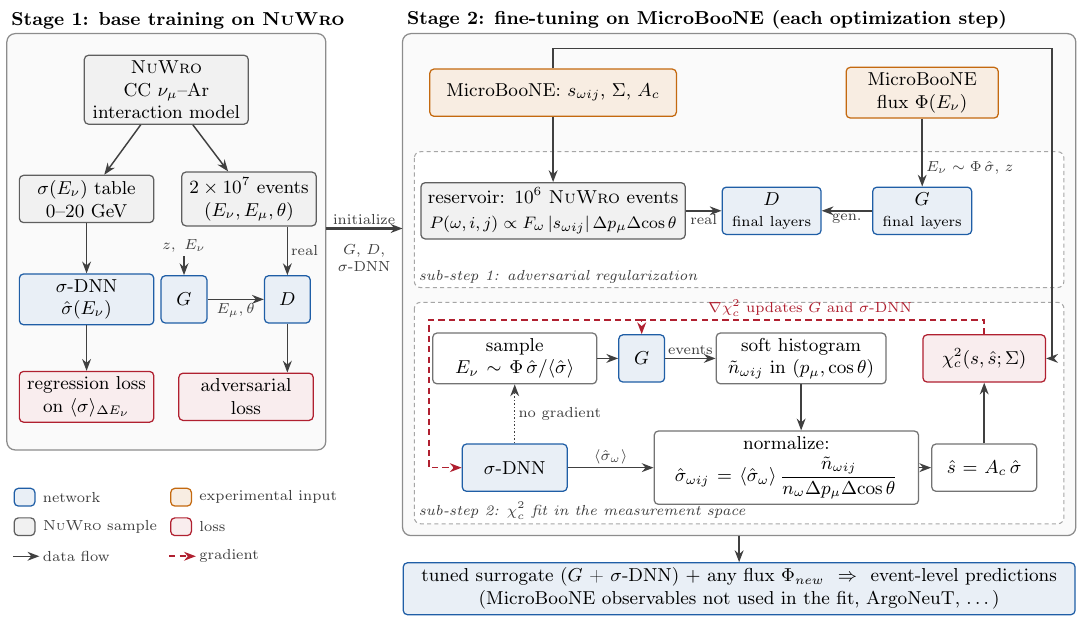}
\caption{Training scheme. (a) The $\sigma$-DNN and the conditional GAN
($G$, $D$) are first trained on NuWro. (b) Fine-tuning combines
(1) adversarial regularization using the data-informed reservoir and
(2) a differentiable full-covariance fit in the measurement space.
The discriminator is updated on every reservoir batch, while every
fifth discriminator update is followed by an adversarial update of
$G$ and by the $\chi_c^2$ update of $G$ and the $\sigma$-DNN.
Sampling of $E_\nu$ from $\Phi\hat\sigma$ carries no gradient. \label{fig:scheme}}
\end{figure*}

The $\sigma$-DNN is a feed-forward network that supplies the interaction cross
section used to weight the events generated by the GAN. Beyond a prediction at a
fixed neutrino energy, it must return the cross section averaged over an energy
interval, so that flux-averaged cross sections can be evaluated over the
measurement bins. We describe in turn the tabulated
training data, the input and target parametrization, the architecture, and the
optimization.

\subsection{Training data and interval averaging}

We first tabulate cross-section values with \nuwro, sampling neutrino energies
from $0$ to $20000$~MeV in steps of $5$~MeV,
\begin{equation}
    E_{\nu,i} \in \{0, 5, \dots, 19995, 20000\}~\text{(MeV)},
\end{equation}
with the corresponding cross sections
\begin{equation}
    \sigma_i \in \{\sigma(0), \sigma(5), \dots, \sigma(19995), \sigma(20000)\}
\end{equation}
in units of $\mathrm{cm}^2$.
Below the reaction threshold, the cross section is set to zero.

To obtain the average cross section over an energy interval $\Delta E_\nu$ (a bin
width, which in our analysis reads $50$~MeV), two table entries with indices $m,n\in\mathbb{N}$, $n\leq m$,
corresponding to $E_{\nu,n}$, $E_{\nu,m}$ and $\sigma_n$, $\sigma_m$, are
selected at random, and the cross section is averaged over all tabulated values
within the interval,
\begin{equation}
    \langle\sigma_{mn}\rangle = \frac{1}{m-n+1}\sum_{i=n}^{m}\sigma_i,
    \label{eq:sigmabin}
\end{equation}
which is the flux-averaged cross section under the assumption of a uniform flux
across the interval; this local-uniformity assumption is what motivates the
finer flux subdivision $\Phi\to\Phi'$ used in following sections. The single-energy
prediction is recovered as the special case $n=m$.

\subsection{Input and target parametrization}

Since the cross section is unbounded, growing approximately linearly at large
neutrino energies, the network is trained not on $\langle\sigma_{mn}\rangle$
directly but on the bounded target
\begin{equation}
    \sigma'_{mn} = \sqrt[4]{\frac{\langle\sigma_{mn}\rangle\times 10^{38}}{E_{mn}}},
    \qquad
    E_{mn} = \frac{(E_{\nu,m} + E_{\nu,n})}{2}\times 10^{-3}, 
    \label{eq:sigmatransform}
\end{equation}
where dividing by $E_{mn}$ bounds the quantity from above, and after rescaling $E_{mn}$ is in GeV units. The physical cross
section is recovered by inverting Eq.~\ref{eq:sigmatransform} after prediction.
The energy inputs are rescaled to the unit interval,
\begin{equation}
    E'_{\nu,m} = \frac{E_{\nu,m}}{2\times 10^{4}}, \qquad E'_\nu \in [0,1],
\end{equation}
and the input for each sample is the tuple
\begin{equation}
    \text{input} = \left(E'_{\nu,m},\; E'_{\nu,n},\; E'_{\nu,m} - E'_{\nu,n}\right).
\end{equation}
To improve convergence, the interval width is constrained during training by
$E_{\nu,m} - E_{\nu,n} \leq 1000$~MeV, so the model is trained only on intervals
narrower than $1$~GeV.

\subsection{Architecture}

The network is built from seven blocks, each consisting of a dense layer
followed by layer normalization and a hyperbolic-tangent activation. The blocks
contain $512$, $256$, $128$, $64$, $32$, $16$, and $8$ neurons, respectively. A
final dense layer produces a single output, activated by a softplus function
\begin{equation}
    f(x) = \frac{\ln\!\left(1 + e^{10x}\right)}{10},
    \label{eq:softplus}
\end{equation}
whose lower bound guarantees a non-negative predicted cross section; the factor
of $10$ controls the sharpness of the softplus knee.

\subsection{Optimization}

Three choices act together to keep the training target $\sigma'_{mn}$ strictly
positive. First, the softplus output cannot be negative. Second, to avoid 
zeros in the tabulated cross sections, we apply
\begin{equation}
    \langle\sigma_{mn}\rangle \rightarrow \langle\sigma_{mn}\rangle + \epsilon,
    \quad \text{if } \langle\sigma_{mn}\rangle = 0,
\end{equation}
with $\epsilon = 14\times10^{-48}$~cm$^2$. Third, the loss is normalized by the target itself,
\begin{equation}
    L = \frac{1}{N}\sum_{mn\,\in\,\text{batch}}
        \frac{\left(\sigma'_{mn} - \hat{\sigma}'_{mn}\right)^2}{\sigma'_{mn}},
    \label{eq:sigmaloss}
\end{equation}
where $\hat{\sigma}'_{mn}$ is the network prediction. This target-normalized
form prevents the large-cross-section region from dominating the loss, so that
the low-energy, near-threshold region, where $\sigma$ is small, remains
well fit; it also requires $\sigma'_{mn} > 0$, which the two preceding steps
ensure. We train with the Nadam optimizer~\cite{dozat2016nesterov} at a learning
rate of $0.001$, using mini-batches of size $N = 1000$. The single-energy result
($m=n$) is shown in Fig.~\ref{fig:base_xsec}.

\begin{figure}[htbp]
    \includegraphics[width=0.48\textwidth]{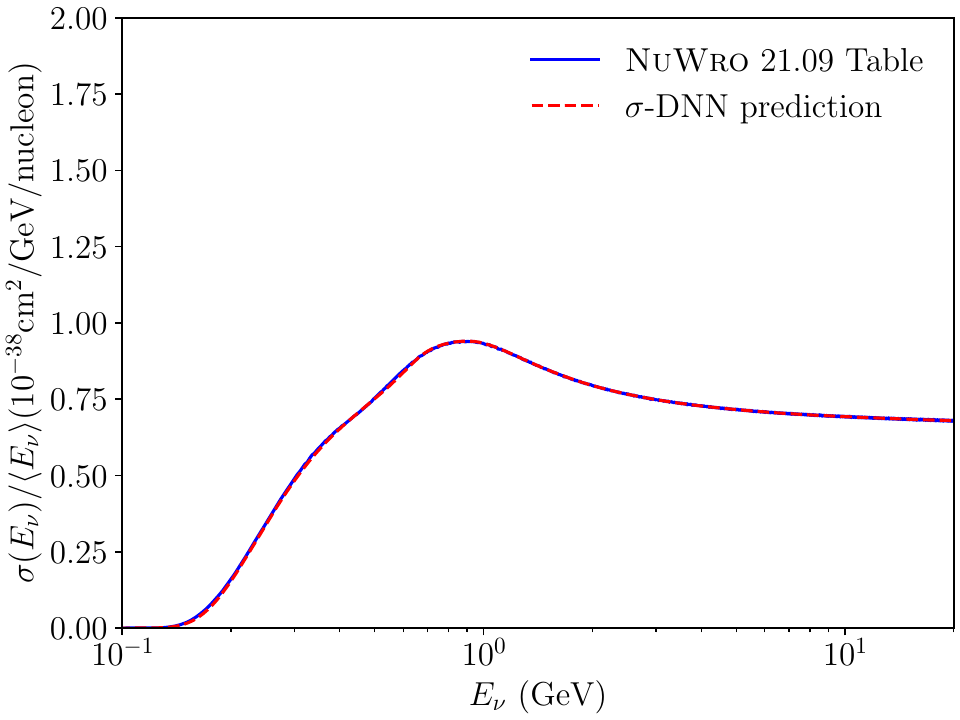}
    \caption{Cross section predicted by \nuwro\ and by the base neural network. Predictions were made on the energies sampled for training (\nuwro) and points in the middle of the energies of the tables (DNN).\label{fig:base_xsec}}
\end{figure}

The trained \(\sigma\)-DNN is subsequently used as a flux reweighter. For each flux-energy bin it provides the cross-section average defined in Eq. (\ref{eq:sigmabin}), which determines the event-energy sampling probability used in the main text. Fig.~\ref{fig:base_xsec} validates the resulting single-energy prediction against the \nuwro{} table.

\section{Generative adversarial network: architecture and training}
\label{app:gan}

A generative adversarial network (GAN) consists of two networks trained in
competition: a generator $G$ and a discriminator $D$. The discriminator learns
to distinguish real samples, drawn from the training set, from fake samples
produced by $G$, while $G$ learns to produce samples that $D$ judges to be real.
At convergence, $G$ produces samples indistinguishable from the real ones. GANs
for $\nu_\mu$ scattering and their domain adaptation via transfer learning were
discussed in Refs.~\cite{bonilla2025generativeadversarialneuralnetworks,%
Bonilla:2025rrn}. We use a conditional GAN (cGAN), since we reproduce the event
kinematics as a function of the neutrino energy $E_\nu$.

\subsection{Implementation and building blocks}
% I haec added diagrams showing the actual architecture. The files are "generator.pdf" and "discriminator.pdf"
We implement the network in Keras~2.15~\cite{chollet2015keras}. The basic
feature-extraction block combines a dense layer, layer
normalization~\cite{ba2016layernormalization}, an activation, and a dropout
layer. The activations are \texttt{ReLU}~\cite{pmlr-v15-glorot11a} for $G$ and
\texttt{PReLU}~\cite{2015arXiv150201852H} for $D$. Zero-centered Gaussian noise
with learnable parameters is injected into the inputs and after each layer (except for the first dense layer that receives the noisy inputs), at a rate of $0.0001$
for $G$ and $0.001$ for $D$.

The generator receives a $50$-dimensional latent vector drawn from
$\mathcal{N}(0,1)$ together with the rescaled neutrino energy. Each of the two
inputs is processed by its own dense layer with 100 neurons, then concatenated, and dropout is applied, and passed to the
main body of $G$. The generator outputs a pair of values representing the muon
kinematic variables, activated by a hyperbolic tangent that constrains them to
$(-1,1)$. The discriminator receives the tuple $(E'_\mu, \theta')$ together with
$E'_\nu$, from either the training set or a generated sample, which are
concatenated and passed to a dense layer with 282 neurons, then dropout, and finally to the main body of $D$. The main body of $G$ comprises
three consecutive blocks of $100$ neurons each; the body of $D$ comprises six
blocks of $141$ neurons each, so that each $D$ block carries roughly twice the
number of parameters of a $G$ block.

\subsection{Variable parametrization}

The networks do not process the physical variables $E_\nu$, $E_\mu$, $\theta$
directly. Instead, we use the transformations
\begin{equation}
\begin{aligned}
    E'_\nu &= \ln\!\left[\frac{E_\nu\,(\text{GeV}) + e}
                              {E_{\nu,\min}\,(\text{GeV}) + e}\right] - 1, \\[4pt]
    E'_\mu &= 2\sqrt{1-\sqrt{1-\sqrt{1-\sqrt{1-\dfrac{E_\mu-m_\mu}{E_\nu-m_\mu}}}}} - 1, \\[4pt]
    \theta' &= 2\sqrt{1-\sqrt{1-\sqrt{1-\sqrt{1-\theta\,(\text{rad})/\pi}}}} - 1,
\end{aligned}
\label{eq:ganparam}
\end{equation}
with $E_{\nu,\min} = 0.11$~GeV (the lowest neutrino energy in the
training \nuwro\ dataset is $E_\nu \approx 0.112$~GeV) and $e$ is the base of the
natural logarithm. Training on the muon kinematics proved difficult, especially
at low energies $E_\nu < 1$~GeV, which motivates this choice: the
parametrization (empirically tuned to flatten the distribution near the phase-space boundaries) constrains the outputs to physically valid events; the muon can
be neither more energetic than the incoming neutrino nor less energetic than its
rest mass, and it pushes the kinematic distributions away from the edges of the
allowed phase space. As a result, $G$ need not drive its weights to large values
to reproduce the distribution, while $D$ can still resolve its tails.

\subsection{Dense skip connections}

Because adversarial training is inherently unstable, we make heavy use of skip
connections to ease the flow of information through both networks. The output of
each block is made available as an input to every subsequent block, yielding
densely connected networks; the inputs to each block are averaged over their
number, as shown in Fig.~\ref{fig:denseskipconnections}. Such architectures have
been explored extensively, e.g.\ residual networks~\cite{DBLP:journals/corr/%
HeZRS15} and densely connected networks~\cite{DBLP:journals/corr/HuangLW16a}.

\begin{figure}[htbp]
    \includegraphics[width=0.46\textwidth]{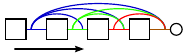}
    \caption{The output of each block of $G$ and $D$ is used as an input to all
    subsequent blocks; all inputs to a block are averaged before it is
    processed.\label{fig:denseskipconnections}}
\end{figure}

\subsection{Training}

The GAN was trained from scratch on a CC $\nu_\mu$--argon sample generated with
\nuwro, to reproduce the muon energy $E'_\mu$ and scattering angle $\theta'$
distributions. The training set contains $20$~million events with neutrino
energies in the range ${\sim}112$--$20000$~MeV. The spectrum is deliberately
irregular, with more events generated at low energies than at high ones, and the
same spectrum is used to feed $G$; this prevents $D$ from learning the energy
spectrum instead of the muon kinematics.

Each discriminator batch consists of $2000$ training events plus $2000$ from
$G$, while each generator batch consists of $4000$ events; one generator step is
taken for every five discriminator steps. The model was trained for several
hundred epochs. The learning rate was initialized at $10^{-4}$ for D and $10^{-5}$ for G and reduced
by a factor of ten in stages over the course of training. The optimizer was
AdamW with \texttt{amsgrad}~\cite{Reddi:2019isk} enabled at the final stages of training.

\section{Fine-tuning}

\subsection{Reservoir construction and the regularization sub-step}
\label{app:reservoir}

The fine-tuning must reconcile two mismatched representations: the GAN operates
on individual events, while the MicroBooNE result~\cite{MicroBooNE:2023foc} is a
three-dimensional differential cross section $d^2\sigma/dp_\mu\,d\cos\theta$ in bins of $E_\nu$. To supply the GAN with event-by-event information informed by the measurement,
we construct a reservoir: a sample of \nuwro\ events whose target
occupancies in $(E_\nu,p_\mu,\cos\theta_\mu)$ are determined from the measured
cross section and the MicroBooNE neutrino flux.

\subsubsection{From the measured cross section to reservoir occupancies}

The MicroBooNE measurement is reported as a smeared cross section $s$, related to a
truth-level prediction $\hat\sigma$ by
\begin{equation}
    \hat s = A_c\,\hat\sigma,
\end{equation}
where $A_c$ is the  smearing matrix provided with the data, acting on
the measurement bins.

Let $s_{\omega_{ij}}$ denote the measured double-differential cross section
in the neutrino-energy region $\omega$, muon-momentum bin $i$, and angular
bin $j$. To construct non-negative sampling probabilities for the reservoir,
we define the bin weight
\begin{equation}
    q_{\omega_{ij}}
    =
    |s_{\omega_{ij}}|\,
    \Delta p_{\mu,i}\,
    \Delta\cos\theta_j .
    \label{eq:reservoir_bin_weight}
\end{equation}
The absolute value is used only in constructing the reservoir because the
regularized unfolding yields negative central values in some low-statistics
bins, which cannot correspond to negative event counts. The original signed
measurement is retained in the subsequent covariance-based fit.

For each neutrino-energy region, we define the corresponding integrated
non-negative weight
\begin{equation}
    Q_\omega = \sum_{ij} q_{\omega_{ij}}.
    \label{eq:reservoir_region_weight}
\end{equation}
Let 
\begin{equation}
F_\omega = \frac{\int_\omega dE_\nu  \Phi(E_\nu)}{\int_{\Omega_\mathrm{res}} dE_\nu  \Phi(E_\nu)}, \qquad \Omega_\mathrm{res} = \bigcup_\omega\omega
\end{equation}
denote the fraction of the MicroBooNE flux within
$\Omega_{\rm res}$ contained in region $\omega$, such that
\[
\sum_\omega F_\omega=1.
\]
The relative population of neutrino-energy regions in the reservoir is then
\begin{equation}
    P(\omega)
    =
    \frac{F_\omega Q_\omega}
    {\displaystyle\sum_{\omega'}F_{\omega'}Q_{\omega'}} .
    \label{eq:reservoir_enu_prob}
\end{equation}
Within a fixed neutrino-energy region, the measured cross section determines
the relative population of the muon-kinematic bins,
\begin{equation}
    P(i,j|\omega)
    =
    \frac{q_{\omega_{ij}}}{Q_\omega}.
    \label{eq:reservoir_cond_prob}
\end{equation}
Consequently, the probability assigned to a three-dimensional reservoir bin is
\begin{equation}
    P(\omega,i,j)
    =
    \frac{
        F_\omega |s_{\omega_{ij}}|
        \Delta p_{\mu,i}\Delta\cos\theta_j
    }{
        \displaystyle
        \sum_{{\omega'}_{i'j'}}
        F_{\omega'} |s_{{\omega'}_{i'j'}}|
        \Delta p_{\mu,i'}\Delta\cos\theta_{j'}
    }.
    \label{eq:reservoir_prob}
\end{equation}
For a reservoir containing $N$ events, the target occupancy of each bin is
therefore $N_{\omega_{ij}}\simeq N P(\omega,i,j)$.

\subsubsection{Drawing the reservoir}

We start from a large \nuwro\ sample generated with the MicroBooNE
neutrino-energy spectrum. For each three-dimensional bin
$(\omega,i,j)$, \nuwro\ events falling in that bin are selected according
to the target occupancy defined by Eq.~(\ref{eq:reservoir_prob}). The
selected events are then combined and shuffled to form the one-million-event
reservoir. The reservoir therefore reproduces the flux-weighted event
occupancies inferred from the measured cross section while retaining
truth-level \nuwro\ kinematics.

The reservoir is not a direct experimental target. In particular, the
measurement used to construct it is smeared, whereas the selected \nuwro\
events retain their truth-level kinematics. The reservoir therefore acts
as a data-informed regularization sample; the quantitative fit to the
measurement is performed separately in the second sub-step.
\begin{figure}[htbp]
    \includegraphics[width=0.48\textwidth]{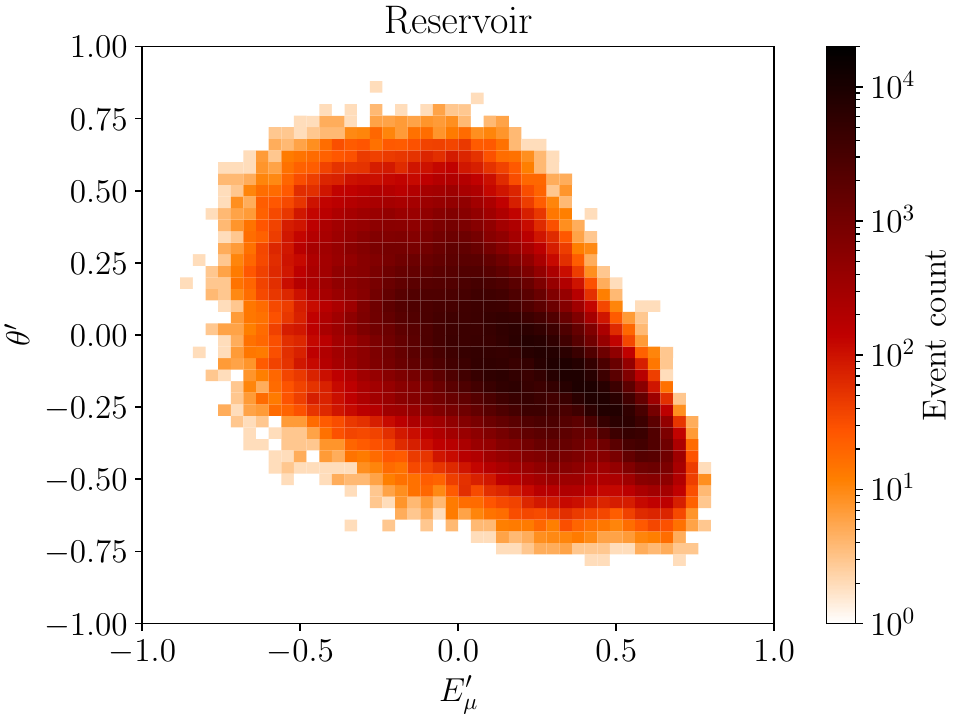}
    
    \caption{One-million-event reservoir used to fine-tune the GAN.
    The reservoir is constructed by resampling truth-level \nuwro\ events
    according to flux-weighted target occupancies inferred from the MicroBooNE
    measurement. The visible bin structures reflect the experimental binning.
    \label{fig:fakedata}}
    
\end{figure}

During base training, the real and generated samples are constructed with the
same $E_\nu$ spectrum, preventing the discriminator from learning to distinguish them from the neutrino-energy distribution alone. During fine-tuning, the generator instead samples $E_\nu$ from the flux reweighted by the current
$\sigma$-DNN prediction. Only the final layers of both the generator and
discriminator are updated, while the preceding layers remain fixed. The $\sigma$-DNN is also partially frozen, with only its final nine
Keras layers left trainable during fine-tuning. This
preserves most of the representation learned during base training while allowing the generated kinematics to adapt to the data-informed reservoir.

During fine-tuning, the discriminator is updated on every reservoir
batch. Every fifth discriminator update is followed by an adversarial
update of the generator and then by the $\chi_c^2$ update of the
generator and the $\sigma$-DNN.

\subsubsection{Role of the reservoir}

An important subtlety concerns what the reservoir represents. Its target
occupancies are derived from the smeared measurement and the
MicroBooNE flux, whereas the reservoir itself consists of truth-level
\nuwro\ events. It therefore cannot be interpreted as a truth-level estimate
of the measurement: the smearing matrix $A_c$ is not inverted at any point,
and such an inversion would in any case be ill-posed given the regularized
unfolding.

In the adversarial sub-step, the reservoir provides the real-event sample
for the update of the discriminator and, through the discriminator, of
the generator.
Unlike the $\chi^2_c$ objective,
the adversarial loss does not compare binned distributions; rather, it
constrains the generated event distribution implicitly through the
discriminator. In this way, the reservoir regularizes the generator toward
a data-informed but physically valid, \nuwro-like manifold of events and
helps prevent mode collapse.

The quantitative comparison with the data is carried out in the second
sub-step: the generated events are histogrammed, normalized using the
$\sigma$-DNN prediction, forward-smeared by $A_c$, and compared with the
original signed measurement through the $\chi^2_c$ objective. Because this
comparison is performed directly in the measurement space, it constrains
both the shape and the normalization.

The two sub-steps thus play complementary roles. Without the first sub-step, the $\chi^2_c$ term alone reproduces the smeared, flux-averaged
distributions but distorts the QE and $\Delta$-resonance structure
(Fig.~\ref{fig:nochi2} of the main text), which the smearing $A_c$ and the
coarse binning render invisible to $\chi^2_c$. The reservoir supplies this
event-level differential structure, while the $\chi^2_c$ term performs the
actual fit to the MicroBooNE data.

\subsection{Second sub-step: differentiable matching to the differential cross section}
\label{app:secondstep}

In the second sub-step the generator $G$ and the $\sigma$-DNN are updated together
so that the generated, cross-section-weighted, forward-smeared distribution matches
the MicroBooNE differential measurement. The central difficulty is that this
matching requires histogramming the generated events, which is not differentiable;
we resolve it with a smooth histogram approximation described below.

\subsubsection{Flux subdivision and the flux-averaged cross section}

The measurement bins $\omega$ are generally wider than, and not aligned with, the
flux bins $\Phi$. We therefore subdivide each $\Phi$ bin into finer sub-bins
$\Phi'$ of width $\Delta E'_\nu$, preserving the flux normalization within each
parent bin, so that every measurement region $\omega$ contains an integer number
of $\Phi'$ bins (Fig.~\ref{fig:fluxbins}). For a region $\omega$ the flux-averaged
cross section can then be obtained in two equivalent ways,
\begin{equation}
    \langle\hat\sigma_\omega\rangle
    = \underbrace{\frac{\sum_{i\in\omega}\hat\sigma_i\Phi'_i}{\sum_{i\in\omega}\Phi'_i}}_\text{prediction}
    = \langle\hat\sigma_\Omega\rangle\,\underbrace{\frac{n_\omega/N}{\sum_{i\in\omega}\Phi'_i}}_\text{MC},
    \label{eq:omegaxsec_app}
\end{equation}
the first using only the $\sigma$-DNN output $\hat\sigma_i$, the second relying on
the sampled events ($n_\omega$ of $N$ falling in $\omega$). The two expressions are analytically equivalent. In the optimization, we use the ``prediction'' expression based directly on the
$\sigma$-DNN output, while the ``MC'' expression makes explicit the
equivalent relation between the flux-averaged cross section and the
fraction $n_\omega/N$ of generated events falling in the region
$\omega$.

\subsubsection{Differentiable histogram}
\begin{figure}[htbp]
  \centering
  \begin{minipage}[b]{0.22\textwidth}
    \vspace{0pt}
    \centering
    \includegraphics[width=\linewidth]{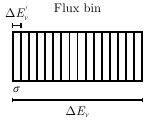}
  \end{minipage}\ \
  \begin{minipage}[b]{0.22\textwidth}
    \vspace{0pt}
    \centering
    \includegraphics[width=\linewidth]{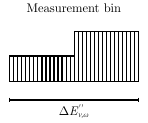}
  \end{minipage}
  \caption{Left: A representation of a flux $\Phi$ bin of width $\Delta E_{\nu}$ divided into sub-bins of flux $\Phi'$ of width $\Delta E'_{\nu}$. Right: Measurement bins of width $\Delta E''_{\nu,\omega}$ that may contain several $\Phi$ bins or fractions, but always entire $\Phi'$ bins.\label{fig:fluxbins}}
\end{figure}

The generator output $(E'_\mu,\theta')$ is first mapped to the physical variables
$(p_\mu,\cos\theta)$. To histogram these events differentiably, we replace the hard
assignment of an event to a bin by a smooth one built from sigmoids \cite{aviaharon2020deephistdifferentiablejointcolor},
$\text{S}(x)=1/(1+e^{-x})$. For a bin with edges $(x_i,x_{i+1})$ and width
$\Delta x$, the contribution of the $n$-th event, with value $x_n$, is the
difference of two sigmoids,
\begin{equation}
    h_i(x_n) = \text{S}\!\left(\alpha\frac{x_n-x_i}{\Delta x}\right)
             - \text{S}\!\left(\alpha\frac{x_n-x_{i+1}}{\Delta x}\right),
    \label{eq:bincontrib_app}
\end{equation}
where $\alpha$ ($=100$) controls the sharpness of the bin edges. The first sigmoid turns on
once $x_n$ passes the left edge, the second once it passes the right edge; their
difference is $\approx 1$ for an event inside the bin and $\approx 0$ outside,
recovering hard binning as $\alpha\to\infty$ while remaining smooth and
differentiable for finite $\alpha$. The soft bin count is
\begin{equation}
    \tilde n_i = \sum_{n=1}^N h_i(x_n),
\end{equation}
and for the two-dimensional $(p_\mu,\cos\theta)$ distribution the joint bin is the
product
\begin{equation}
    h_{ij}(x_n,y_n) = h_i(x_n)\,h_j(y_n), \qquad
    \tilde n_{ij} = \sum_{n=1}^N h_{ij}(x_n,y_n).
\end{equation}
Because $\tilde n_{ij}$ is a smooth function of the event positions, and these
depend on the generator weights, the gradient of the loss propagates through the
histogram back to $G$; a hard histogram would give a vanishing gradient and no
training signal. This is the standard soft-histogram
construction~\cite{aviaharon2020deephistdifferentiablejointcolor}; its accuracy to exact binning at our
chosen $\alpha$ is shown in Fig.~\ref{fig:hists}. The soft count is an estimator of
the exact count; events near a bin edge are shared between neighbors, negligible
at large $\alpha$ but validated in Fig.~\ref{fig:hists}.
\begin{figure}[htbp]
    \includegraphics[width=0.48\textwidth]{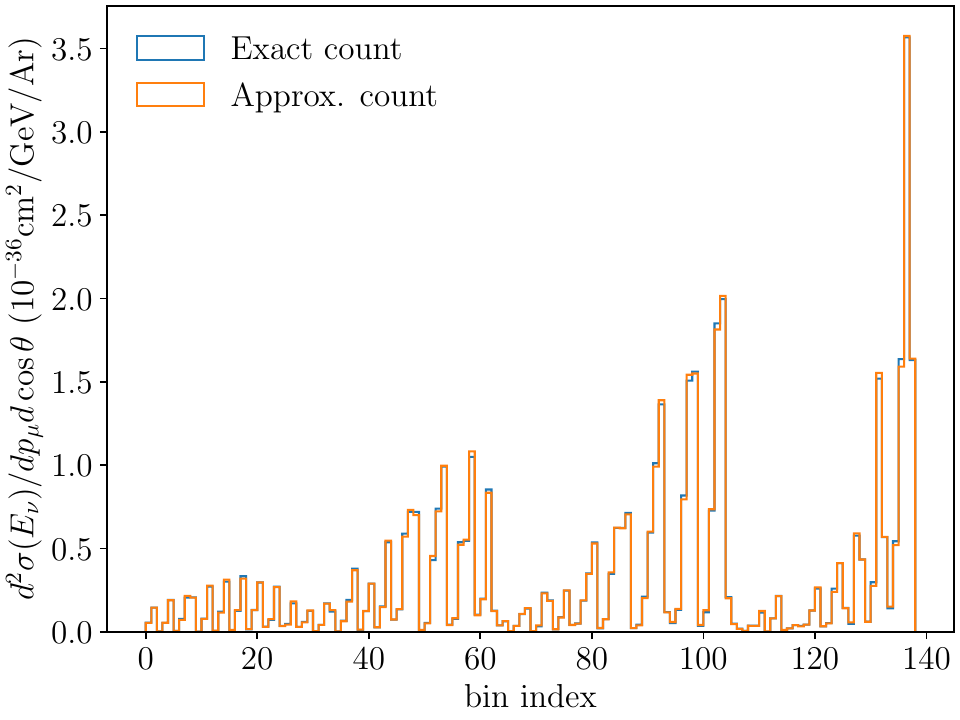}
    \includegraphics[width=0.48\textwidth]{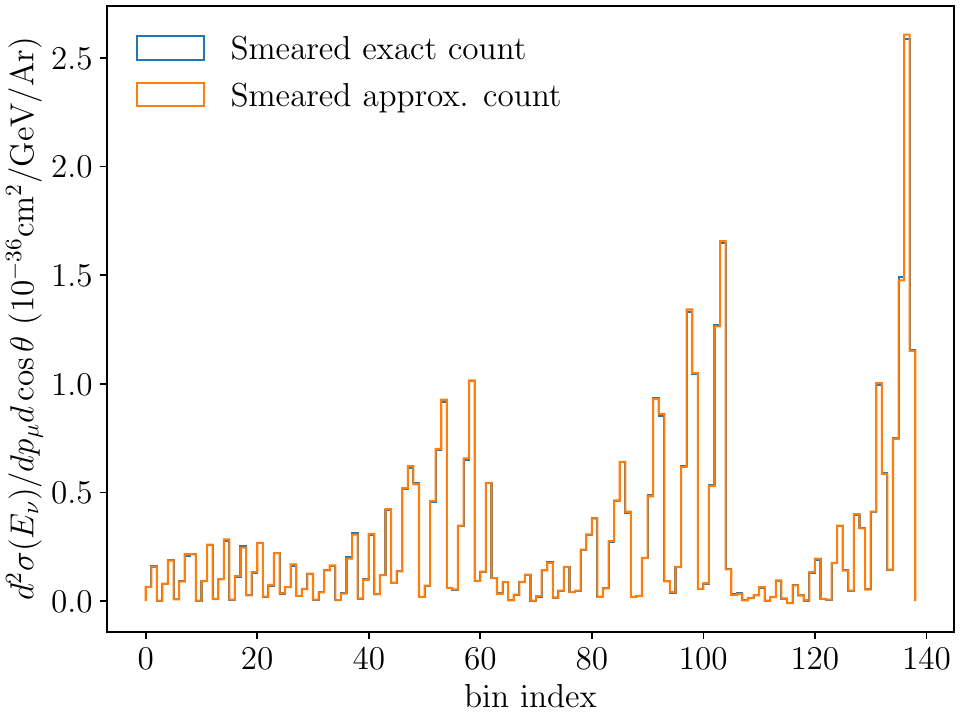}
    \caption{Left: double differential cross section as a function of $E_\nu$ from the base DNN models, using exact event counts to create histograms, and the result using the differentiable histogram approximation, with a batch of 4000 events. Right: same data as left, but smeared with the matrix $A_c$.\label{fig:hists}}
\end{figure}

\subsubsection{From histogram to smeared prediction}

The soft histogram is reweighted by the flux-averaged cross section to form an
estimate of the double-differential cross section in region $\omega$,
\begin{equation}
    \left(\frac{d^2\hat\sigma_\omega}{dx dy}\right)_{ij}
    = \langle\hat\sigma_\omega\rangle\,\frac{\tilde n_{\omega_{ij}}}{n_\omega\,\Delta x_i\,\Delta y_j},
    \label{eq:estimateddiffxsec_app}
\end{equation}
where \(n_\omega\) is the number of generated events with \(E_\nu\) in the energy interval \(\omega\). Eq. (\ref{eq:estimateddiffxsec_app}) is then forward-smeared by the additional smearing matrix,
$\hat s_{\omega,ij} = A_c\,\hat\sigma_{\omega,ij}$, bringing the prediction into the
same space as the published measurement. The smearing is applied to the prediction
and never inverted on the data, so the comparison is well-posed.

\subsubsection{Gradient flow and the role of the $\sigma$-DNN}

The $\sigma$-DNN enters this sub-step in two distinct roles. As the reweighter that
defines the sampling distribution of $E_\nu$, it is treated as a fixed input and
does not contribute to the gradient. As the predictor of the flux-averaged cross
section $\langle\hat\sigma_\omega\rangle$ that rescales the histogram
[Eq.~\ref{eq:estimateddiffxsec_app}], it is differentiated, and this is the
path through which its weights are updated. Thus the sampling of $E_\nu$ carries no
gradient, while the cross-section normalization does.

\subsubsection{Loss}

The generator and $\sigma$-DNN weights are updated by minimizing the
full-covariance $\chi^2$, 
\begin{equation}
    \chi^2_c = \sum_{kl}\left(s_k - \hat s_k\right)\Sigma^{-1}_{kl}\left(s_l - \hat s_l\right),
    \label{eq:chic_app}
\end{equation}
where $s_k$ is the measured smeared cross section, $\hat s_k$ the smeared prediction
of Eq.~\ref{eq:estimateddiffxsec_app}, $\Sigma$ the measurement covariance, and
$k,l$ index the flattened $(\omega,i,j)$ bins.

Additionally, as a complementary diagnostic, we also report the diagonal metric
\begin{equation}
\chi^2_v = \sum_{\omega_{ij}} \frac{\left(s_{\omega_{ij}} - \hat s_{\omega_{ij}}\right)^2}
{\text{Var}_{\omega_{ij}}},
\label{eq:chiv_main}
\end{equation}
where $\text{Var}_{\omega_{ij}}$ is the variance of $s_{\omega_{ij}}$.

\subsection{Training results}

In Fig.~\ref{fig:histstune_app}, we repeat Fig.~\ref{fig:histstune} with additional plots of (MC$-$Data)$/$Uncertainty to show the quality of our model's predictions.

\begin{figure}[htbp]
    \includegraphics[width=0.48\textwidth]{hist_full}
    \includegraphics[width=0.48\textwidth]{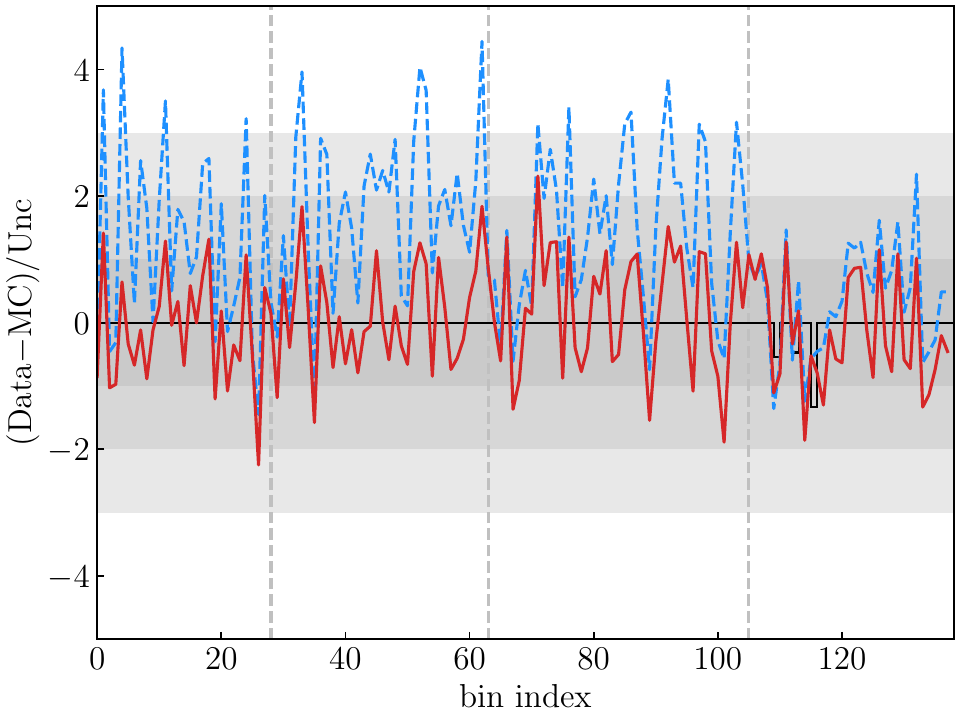}
    \caption{Base and tuned models compared with MicroBooNE two-dimensional differential cross-section data in $ E_\nu$ Bins~\cite {MicroBooNE:2023foc} on the left, and the (MC$-$Data)$/$Unc ratio on the right. The black plot in the ratio figure shows Data$-|$Data$|$ to highlight the negative cross-section values found in the experimental measurement, and the gray bands show the measurement uncertainties. Grey vertical dashed lines show the different $E_\nu$ regions.\label{fig:histstune_app}}
\end{figure}

\section{Comparison with out-of-training MicroBooNE measurements}
\label{app:out-of-training MicroBooNE}

\begin{table*}[htbp]
    \caption{$\chi_c^2$ and $\chi_v^2$ over the number of bins of the base and tuned DNN models for $\sigma(E_\nu)$, $d\sigma/dE_\mu$, and $d\sigma/dE_\text{had}$. \label{tab:chi2xsec}}
    \begin{ruledtabular}
        \begin{tabular}{lcccccc}
                       & \multicolumn{2}{c}{$\sigma(E_\nu)$} & \multicolumn{2}{c}{$d\sigma/dE_\mu$} & \multicolumn{2}{c}{$d\sigma/dE_\text{had}$} \\
                       & $\chi^2_c$/n & $\chi^2_v$/n     & $\chi^2_c$/n & $\chi^2_v$/n      & $\chi^2_c$/n & $\chi^2_v$/n      \\ \hline%\hline
            % chi2_c training
            Base   & 58.1/10        & 54.2/10            & 29.1/11        & 69.1/11             & 17.1/8         & 41.3/8              \\ %\hline
            Tuned & 44.2/10        & 0.5/10             & 12.7/11        & 6.4/11              & 4.4/8          & 6.0/8               \\ %\hline
             \end{tabular}
    \end{ruledtabular}
\end{table*}

In Table~\ref{tab:chi2xsec}, we present the $\chi^2$ values for comparison with MicroBooNE measurements~\cite{MicroBooNE:2021sfa} for $\sigma/\langle E_\nu\rangle$, $d\sigma/dE_\mu$ and  $d\sigma/dE_\text{had}$ cross sections. These numbers refer to Fig.~\ref{fig:sigma_enu}.

In Fig.~\ref{fig:double_app}, we compare our models' predictions for the $d^2\sigma/dE_\mu d\cos\theta$ and $d\sigma/d\cos\theta$ cross-section distributions. Table~\ref{tab:chi2theta} gives the $\chi^2_c$ and $\chi^2_v$ values for the base and tuned models. 

\begin{figure}[htbp]
    \includegraphics[width=0.46\textwidth]{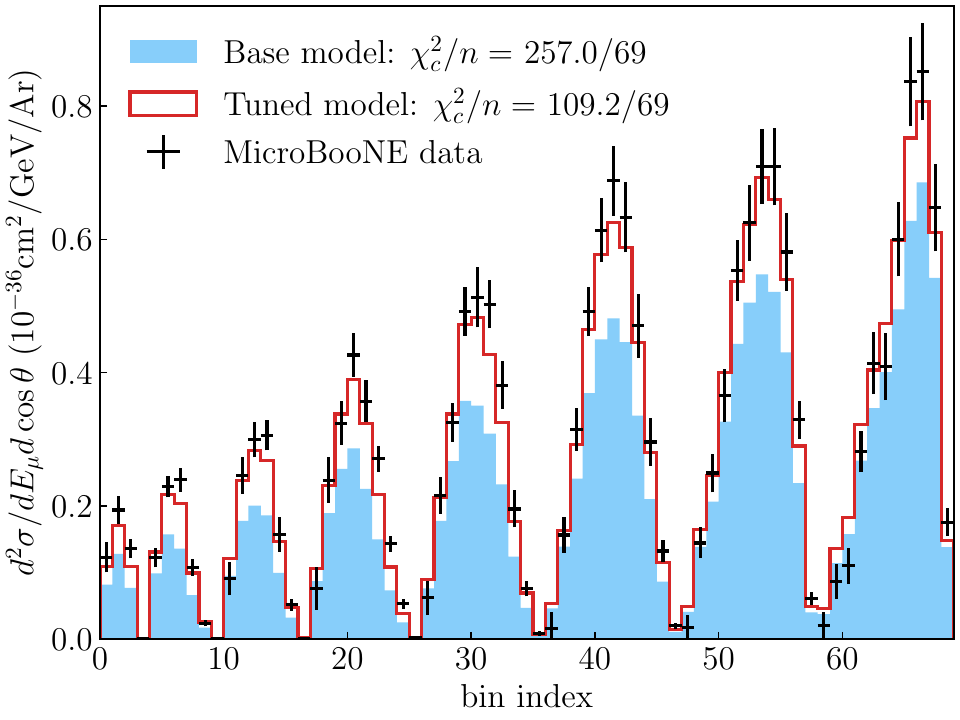} \ \ \
    \includegraphics[width=0.48\textwidth]{costheta.pdf}
    \caption{Base model, tuned model, and MicroBooNE data~\cite{MicroBooNE:2024zkh,MicroBooNE:2024zwf} for $d^2\sigma/dE_\mu d\cos\theta$ (left) and $d\sigma/d\cos\theta$ (right).\label{fig:double_app} 
    }
\end{figure}
\begin{table}[htbp]
    \caption{Same as Table~\ref{tab:chi2xsec} but for $d^2\sigma/dE_\mu d\cos\theta$ and $d\sigma/d\cos\theta$. \label{tab:chi2theta}}
    \begin{ruledtabular}
        \begin{tabular}{lcccc}
                       & \multicolumn{2}{c}{$d^2\sigma/dE_\mu d\cos\theta$}  & \multicolumn{2}{c}{$d\sigma/d\cos\theta$} \\
                       & $\chi^2_c$/n   & $\chi^2_v$/n                       & $\chi^2_c$/n   & $\chi^2_v$/n   \\ \hline%\hline
            Base  & 257.0/69       & 554.7/69                           & 291.5/17       & 2991.7/17      \\ %\hline
            Tuned & 109.2/69       & 87.9/69                            & 6.8/17         & 4.8/17         \\ %\hline
        \end{tabular}
    \end{ruledtabular}
\end{table}

\section{ArgoNeuT results for neutrinos in neutrino mode.}

In Fig.~\ref{fig:numi_app} we present both distributions of cross sections reported by ArgoNeuT~\cite{ArgoNeuT:2011bms, ArgoNeuT:2014rlj}, namely $d\sigma/d\theta$ and $d\sigma/dp_\mu$. 
Table~\ref{tab:chi2argoneutnuonnu} compares the base and tuned models to the
ArgoNeuT $d\sigma/d\theta$ and $d\sigma/dp_\mu$ measurements.

\begin{figure}[htbp]
    \includegraphics[width=0.48\textwidth]{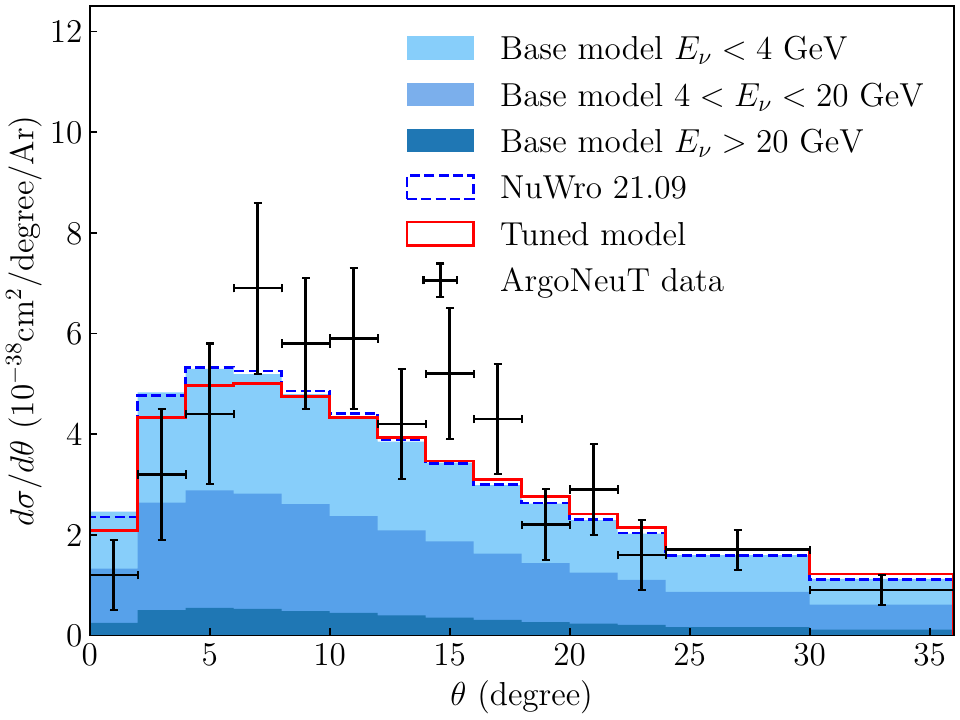} \ \ \
    \includegraphics[width=0.48\textwidth]{nu_on_nu_numi_pmu_nuw.pdf}
    \caption{Base model, tuned model, and ArgoNeuT data~\cite{ArgoNeuT:2011bms, ArgoNeuT:2014rlj}. Left: $d\sigma/d\theta$. Right: $d\sigma/dp_\mu$. 
    Results for neutrinos in the NuMI neutrino beam. The shaded areas on the right-side plot show the $p_\mu$ range of the training MicroBooNE data.\label{fig:numi_app}}
\end{figure}
\begin{table}[htbp]
    \caption{Comparison of MicroBooNE-tuned DNN to ArgoNeuT measurement of muon neutrino $\nu_\mu$ on NuMI neutrino beam.\label{tab:chi2argoneutnuonnu}}
    \begin{ruledtabular}
        \begin{tabular}{lcc}
                       & \multicolumn{2}{c}{$\chi^2_v$/n}   \\
                       & $d\sigma/d\theta$ & $d\sigma/dp_\mu$ \\ \hline%\hline
            Base  & 13.2/14           & 6.1/11           \\ %\hline
            Tuned & 11.4/14           & 6.0/11           \\ %\hline
        \end{tabular}
    \end{ruledtabular}
\end{table}

\section{ArgoNeuT results for neutrinos in antineutrino mode}
\label{app:anineutrino_nutev}

In this section, we compare predictions from our system with ArgoNeuT measurements for neutrinos in an antineutrino beam. Table~\ref{tab:chi2argoneutnuonnu_anti} 
presents the $\chi^2_v$ values computed for fine-tuned and base models. In Fig.~\ref{fig:numi_anti} we show the distributions of $d\sigma/d\theta$ and $d\sigma/dp_\mu$ cross sections and predictions of fine-tuned and base models. 

\begin{table}[htbp]
    \caption{Comparison of MicroBooNE-tuned DNN to ArgoNeuT measurement of muon neutrino $\nu_\mu$ on NuMI anti-neutrino $\bar{\nu}_\mu$ beam.\label{tab:chi2argoneutnuonnu_anti}}
    \begin{ruledtabular}
        \begin{tabular}{lcc}
                       & \multicolumn{2}{c}{$\chi^2_v$/n}   \\
                       & $d\sigma/d\theta$ & $d\sigma/dp_\mu$ \\ \hline%\hline
            Base   & 9.1/18            & 9.6/20           \\ %\hline
            Tuned  & 9.6/18            & 9.1/20           \\ %\hline
        \end{tabular}
    \end{ruledtabular}
\end{table}

\begin{figure}[htbp]
    \includegraphics[width=0.48\textwidth]{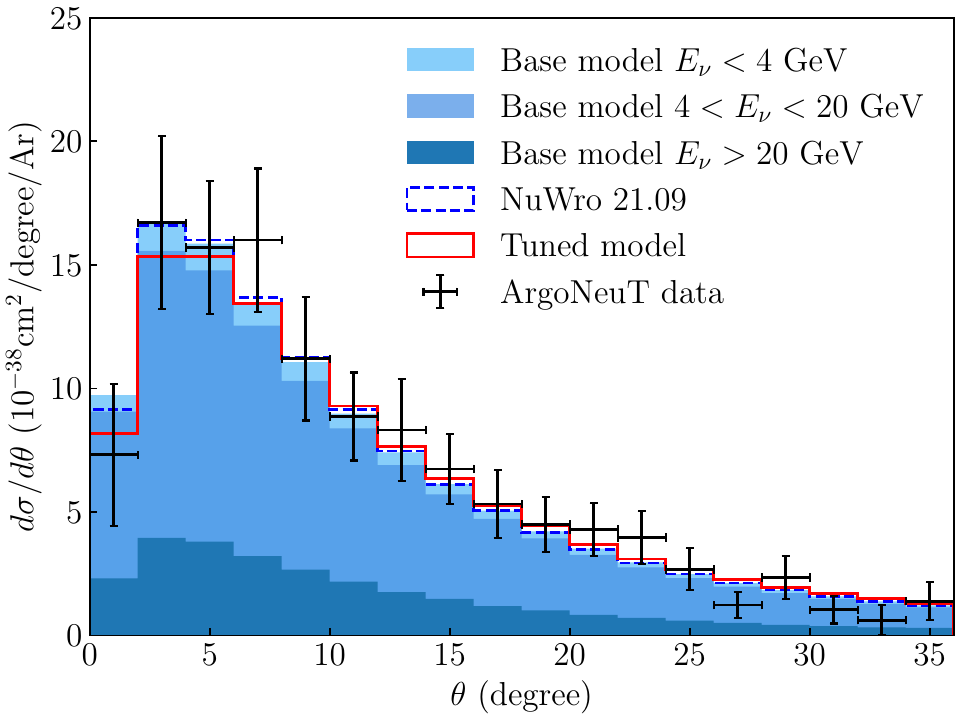}
    \includegraphics[width=0.48\textwidth]{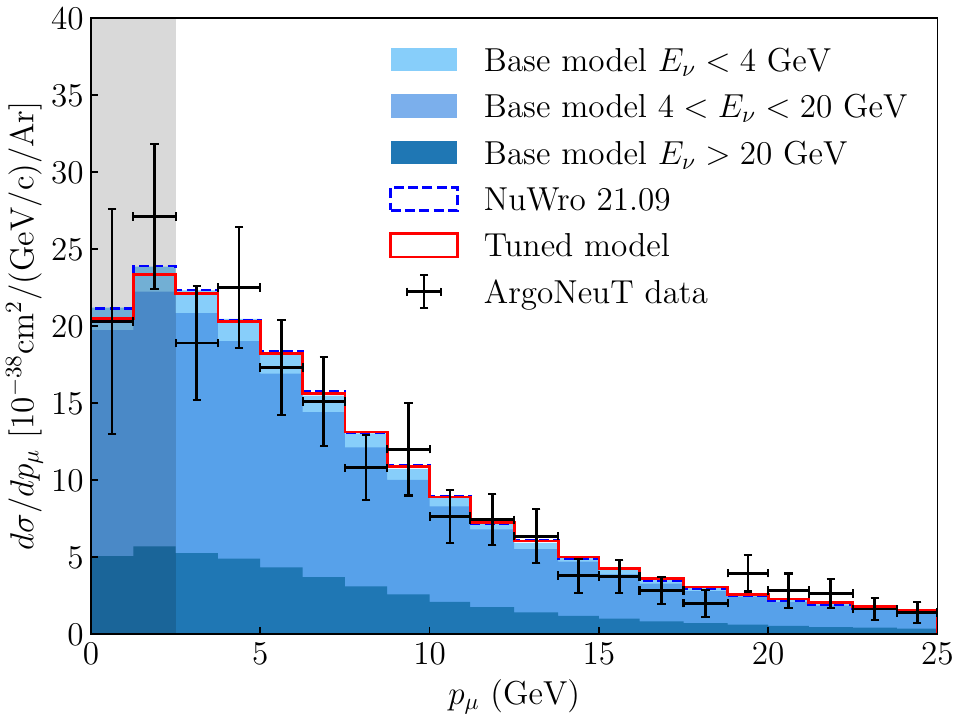}
    \caption{Base model, tuned model, and ArgoNeuT data~\cite{ArgoNeuT:2011bms, ArgoNeuT:2014rlj}. Left: $d\sigma/d\theta$. Right: $d\sigma/dp_\mu$. Results for neutrinos on the NuMI antineutrino beam. The shaded areas on the right-side plots show the $p_\mu$ range of the training MicroBooNE data.\label{fig:numi_anti}}
\end{figure}

\fi

\bibliographystyle{apsrev4-2}
\bibliography{dnn,bibmoje, physics,bibdata}

\end{document}